\documentclass[amssymb,prd,preprint,aps,nofootinbib]{revtex4}
\usepackage{amsmath,amssymb}
\usepackage{graphicx}
\usepackage{epsfig}
\usepackage{epstopdf}
\usepackage{tikz}

\def\bea{\begin{eqnarray}}
\def\eea{\end{eqnarray}}
\begin{document}
\preprint{
CTPU-14-04
}
\title{Natural inflation with multiple sub-Planckian axions}
\author{Kiwoon Choi,\,$^{a,}$\footnote{kchoi@kaist.ac.kr} Hyungjin Kim,\,$^{a,b,}$\footnote{hjkim06@kaist.ac.kr} and Seokhoon Yun$^{a,b,}$\footnote{ yunsuhak@kaist.ac.kr}}
\affiliation{
	$^a$Center for Theoretical Physics of the Universe, Institute for Basic Science (IBS), Daejeon 305-811, Korea \\
	$^b$ 
	Department of Phyiscs, KAIST, Daejeon 305-701, Korea
}
\begin{abstract}
We extend the Kim-Nilles-Peloso (KNP) alignment mechanism for natural inflation to models with $N>2$ axions, which obtains  super-Planckian {effective} axion decay constant $f_{\textrm{eff}}\gg M_{Pl}$ through an alignment of the anomaly coefficients of multiple axions having sub-Planckian fundamental decay constants $f_0\ll M_{Pl}$. The original version of the KNP mechanism realized with two axions requires that some of the anomaly coefficients should be of the order of $f_{\textrm{eff}}/f_0$, which would be uncomfortably large if $f_{\rm eff}/f_0 \gtrsim {\cal O}(100)$ as suggested by the recent BICEP2 results.    We note that the KNP mechanism can be realized with the anomaly coefficients of $\mathcal{O}(1)$ if the number of axions $N$ is large as $N\ln N\gtrsim 2\ln (f_{\textrm{eff}}/f_0)$, in which case the effective decay constant can be enhanced as $f_{\rm eff}/f_0 \sim \sqrt{N !}\,n^{N-1}$ for   $n$ denoting the typical size of the integer-valued anomaly coefficients.
	 Comparing to the other multiple axion scenario, the $N$-flation scenario which requires $N \sim f_{\textrm{eff}}^2/f_0^2$, the KNP mechanism has a virtue of not invoking to a too large number of axions, although it requires a specific alignment of the anomaly coefficients,
which can be achieved with a probability of ${\cal O}(f_0/f_{\rm eff})$  under a random choice of the anomaly coefficients.  We also present  a simple model realizing a multiple axion monodromy  along the inflaton direction.

%The same mechanism can be applied for a quintessence axion which needs a super-Planckina axion decay %constant.	 

\end{abstract}

\maketitle
\section{Introduction}\label{sec:1}

Recent detection of tensor modes in the cosmic microwave background by BICEP2 suggests that the inflation scale is rather high, $H_I\sim 10^{14}$ GeV \cite{Ade:2014xna}.  In the context of slow roll inflation, such a high inflation scale implies that the inflaton field has experienced a super-Planckian excursion 
over the inflation period~\cite{Lyth:1996im}. This poses a question 
if the inflaton field can be decoupled from high scale physics above the scale of quantum gravity, so that an effective field theory description of inflation does make sense.   

An attractive solution to this puzzle is to introduce an   approximate continuous shift symmetry along the inflaton direction.
To implement this idea, in natural inflation~\cite{Freese:1990rb} the inflaton is assumed to be a pseudo-Nambu-Goldstone boson $\phi$  with a periodic potential
\bea
V(\phi)\,=\, \Lambda^4\left[1-\cos\left(\frac{\phi}{f}\right)\right],\eea
where $f$ is the axion decay constant which defines the fundamental domain of the  axionlike inflaton field:
\bea
\phi \,\equiv\, \phi + 2\pi f.\eea
This inflaton potential is stable against perturbative quantum corrections, which can be assured through the  approximate shift symmetry 
$\phi\,\rightarrow \, \phi+\mbox{constant}$.
Also, under a reasonable assumption on the nonperturbative dynamics generating the axion potential, 
one can justify that the above simple potential provides a good approximation to the full inflaton potential.

However, there is a difficulty in this simple setup. For  successful slow roll inflation,
the axion decay constant is required to have a super-Planckian value.
The recent BICEP2 result makes the problem even more severe as it suggests~\cite{Freese:2014nla} \bea
f \,\gtrsim \, 10 M_{Pl},\eea
where $M_{Pl}\simeq2.4\times 10^{18}$ GeV is the reduced Planck mass.
It appears to be difficult to get such a large axion decay constant from a sensible fundamental theory, particularly from string  theory. In the limit where a controllable approximation is available, string theory predicts that the axion scale is typically around $g^2M_{Pl}/8\pi^2$~\cite{Choi:1985je}.

During the past years, there have been several  proposals to circumvent this problem.
They include the two-axion model of Kim {\it et al}.~\cite{Kim:2004rp},
which obtains a super-Planckian {effective} axion decay constant through an alignment of the anomaly coefficients of two axions having sub-Planckian fundamental decay constants, 
a five-dimensional (5D) gauge-axion unification model in which the axionlike inflaton obtains a super-Planckian decay constant in the limit that 5D gauge coupling becomes weaker than the 5D gravitational coupling~\cite{ArkaniHamed:2003wu},  a model with non-minimal axion-gravity coupling where gravitationally enhanced Hubble friction makes natural inflation operative with sub-Planckian axion decay constant~\cite{Germani:2010hd}, the $N$-flation scenario~\cite{Dimopoulos:2005ac} based on the idea of assisted inflation~\cite{Liddle:1998jc} with many axions, and the axion monodromy based on either a string theoretic~\cite{Silverstein:2008sg,Marchesano:2014mla} or field theoretic \cite{Kaloper:2008fb,Harigaya:2014eta} scheme for multiple windings in the axion field space. 
 In this paper, we revisit the Kim-Nilles-Peloso (KNP) alignment mechanism to extend the scheme to  models with $N>2$ axions.
 
  The original version of the KNP mechanism realized with two axions requires that some of the anomaly coefficients should be of the order of $f_{\textrm{eff}}/f_i$, where $f_{\rm eff}\gg M_{Pl}$ is the super-Planckian effective decay constant of the axionic inflaton, while $f_i\ll M_{Pl}$ are the sub-Planckian fundamental axion decay constants in the model.
 In case that $f_i \sim g^2 M_{Pl}/8\pi^2$~\cite{Choi:1985je} as suggested by string theory, this would require that
 some anomaly coefficients should be uncomfortably large as $f_{\rm eff}/f_i = {\cal O}(10^2 -10^3)$.    We note that the KNP mechanism can be realized with the anomaly coefficients of $\mathcal{O}(1)$ if the number of axions is large as $N \ln N \gtrsim 2\ln (f_{\textrm{eff}}/f_i)$, in which case the effective decay constant can be enhanced as $f_{\rm eff}/f_i \sim \sqrt{N!} \, n^{N-1}$ for   $n$ denoting the typical size of the integer-valued anomaly coefficients.
We examine also 
the probability for the KNP alignment to be achieved under a random choice of the anomaly coefficients.

In regard to enhancing the effective axion decay constant, a relevant question  is how many fields do we need  to get  super-Planckian $f_{\rm eff} \gg M_{Pl}$. As the Planck scale receives a quadratically divergent radiative correction from each light field,  schematically we have $\delta M_{Pl}^2
\propto N_l \Lambda^2$, where $N_l$ denotes the number of light fields and $\Lambda$ is the cutoff scale of loop momenta. Then the scheme would be in trouble if it requires a too large number of light fields as
$N_l \geq f^2_{\rm eff}/f^2_i$.  
		In our multiple axion scenario,  $f_{\rm eff}/f_i$ grows exponentially as a function of $N$ for a fixed value of $n>1$, so the number of required axions is of the order of $\ln (f_{\rm eff}/f_i)$.
		On the other hand, for the original two-axion KNP model~\cite{Kim:2004rp}, one needs $n={\cal O}(f_{\rm eff}/f_i)$,
		where the anomaly coefficient $n$ can be identified as the number of gauge-charged fermions generating the  axion coupling to instantons. 	In the $N$-flation scenario~\cite{Dimopoulos:2005ac}, the number of required axions is  ${\cal O}(f^2_{\rm eff}/f^2_i)$.
		  So our scheme can enhance $f_{\rm eff}$  by introducing a parametrically smaller number of fields, as compared to the KNP two-axion model and the $N$-flation scenario.

In certain cases, the KNP mechanism can be interpreted as enhancing the effective axion decay constant as $f_{\rm eff}/f_i \sim n\gg  1$ through the  $Z_n$  monodromy structure of a light axion, which is induced along the inflaton direction by the  mixing with heavy axions.   
In this context, we present a simple model yielding $f_{\rm eff}/f_i \sim \prod_{i=2}^N n_i$ through
 a multiple axion monodromy  described by  $\prod_{i=2}^N Z_{n_i}$ ($n_i>1$).
We present also a model yielding $f_{\rm eff}/f_i \sim 2^{N-1}$ even when all the integer-valued anomaly coefficients are restricted as  $|n_{ij}|\leq 1$.

The organization of this paper is as follows.
In Sec. \ref{sec:2}, we review the original two-axion model of KNP to illustrate the basic idea and set the notations. In Sec. \ref{sec:3}, we extend the KNP mechanism to models with $N>2$ axions.
Sec. \ref{sec:4} is the conclusion.

\section{Kim-Nilles-Peloso mechanism with two axions}\label{sec:2}

We begin with a brief review of the original Kim-Nilles-Peloso mechanism realized with two axions
\cite{Kim:2004rp}.
In the field basis of {\it periodic} axions: 
\bea 
\phi_i \equiv \phi_i + 2\pi f_i,\eea
the axion potential consistent with the axion periodicity is generically given by
\begin{eqnarray}
	V(\phi_i)=\Lambda_1^4\left[1-\cos\left(\frac{n_1\phi_1}{f_1}+\frac{n_2\phi_2}{f_2}\right)\right]+\Lambda_2^4\left[1-\cos\left(\frac{m_1\phi_1}{f_1}+\frac{m_2\phi_2}{f_2}\right)\right],
	\label{potential.2axion}
\end{eqnarray}
where $\vec{n}=(n_1, n_2)$ and $\vec{m}=(m_1, m_2)$ are linearly independent integer-valued  coefficients,
 and $f_i$ ($i=1,2$) denote the fundamental axion decay constants which are presumed to be 
 comparable to each other, while being significantly lower than the reduced Planck scale:
$$ f_1\,\sim\, f_2 \,\ll\, M_{Pl}.$$
Here we include only the leading nonperturbative effects generating the axion
potential, under the assumption  that the next order nonperturbative effects are small enough.

The integer-valued coefficients $n_i, m_i$ parametrize the discrete degrees of freedom in the underlying nonperturbative dynamics generating the axion potential.  A simple possibility is that the axion potential is generated by hidden sector gauge field instantons through the symmetry breaking by anomalies.
In such case,
the model involves two non-Abelian hidden sector gauge groups $G_a$ ($a=1,2$), together with the gauge-charged fermions having 
the following couplings to axions:
\bea
\sum_I \sum_i \lambda_{iI} f_i e^{iq_{iI}\phi_i/f_i} \bar\psi_{IL}\psi_{IR}+{\rm H.c.},\eea
where $\lambda_{iI}$ denote dimensionless Yukawa couplings and  $\psi_I$ are assumed to be charged Dirac fermions for simplicity.
Then     
the Noether current of the nonlinearly realized Peccei-Quinn symmetries \bea 
U(1)_i: \, \phi_i\rightarrow \phi_i + \alpha_i f_i, \quad \bar\psi_{IL}\psi_{IR}\rightarrow e^{-iq_{iI}\alpha_i}\bar\psi_{IL}\psi_{IR} \eea
have the $U(1)_i$-$G_a$-$G_a$ anomalies as 
\bea
\partial_\mu J^\mu_i = \frac{n_i}{16\pi^2}F_1\tilde F_1 + \frac{m_i}{16\pi^2}F_2\tilde F_2,\eea
where $F_a$ are the gauge field strength of the gauge group $G_a$, and the anomaly coefficients are given by
\bea
n_i = 2\sum_I q_{iI} {\rm Tr}(T_1^2(\psi_I)), \quad
m_i = 2\sum_I q_{iI} {\rm Tr}(T_2^2(\psi_I))\eea
for $T_a(\psi_I)$ ($a=1,2$) denoting the $G_a$-charge matrix of $\psi_I$ normalized as
${\rm Tr}(T_a^2)=1/2$ for the fundamental representation of $G_a$. With this symmetry breaking by anomalies, 
the gauge field instantons of $G_a$  generate the axion potential of the form (\ref{potential.2axion}).
Based on this observation, in the following we will call
$n_i,m_i$ the anomaly   coefficients. However it should be noted that  the axion potential (\ref{potential.2axion}) can be generated by different  kinds of nonperturbative effects, for instance string theoretic instantons or hidden gaugino condensations. In such case, the integer
coefficients $n_i, m_i$ can be determined by a variety of different discrete UV quantum numbers, e.g.
the quantized fluxes, the number of stacked $D$-branes, and/or the number of windings for stringy instantons. 

\begin{figure}
\begin{minipage}{6cm}
\includegraphics[scale=0.8]{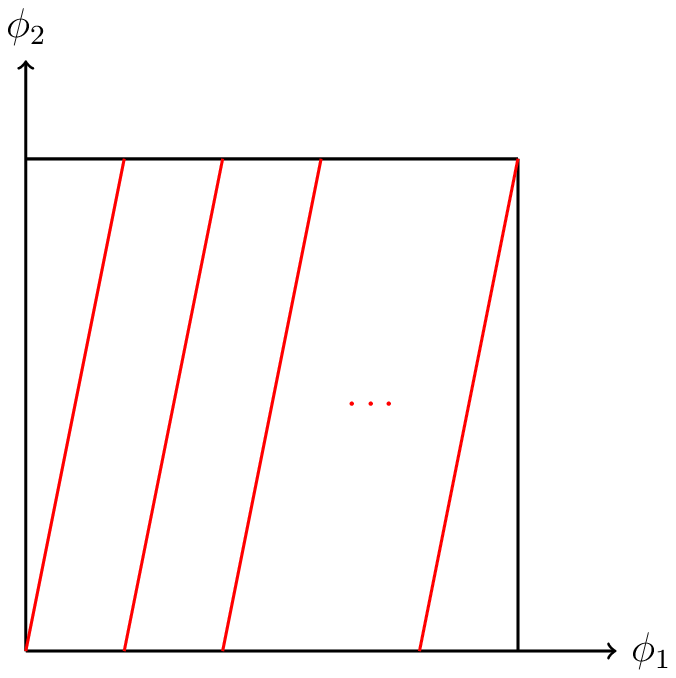}
\end{minipage}
\hspace*{1cm}
\begin{minipage}{6cm}
\vspace{1cm}
\includegraphics[scale=0.6]{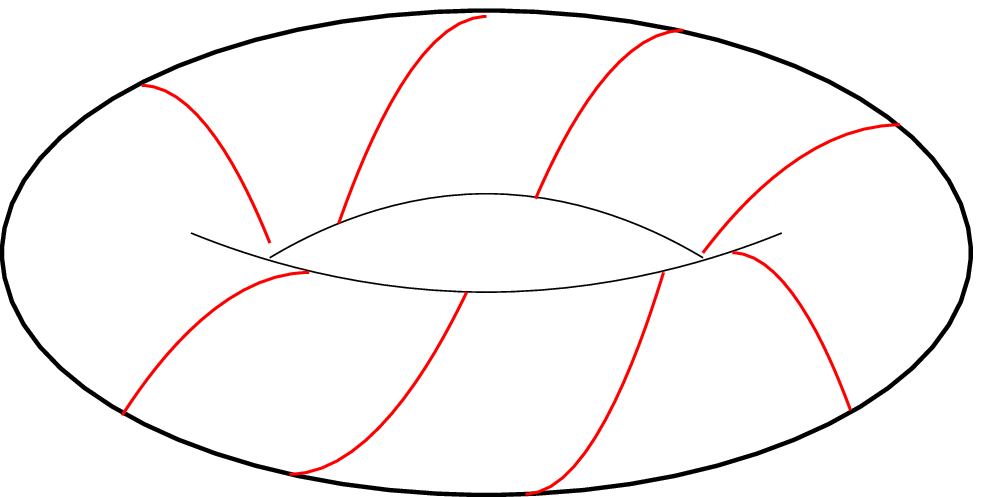}
\end{minipage}
\caption{
Flat direction in the fundamental domain of axion fields in the limit $\Lambda_2=0$.   Even though the fundamental domain is sub-Planckian with $f_i\ll M_{Pl}$, the flat direction can have a super-Planckian length if one (or both) of $n_i/{\rm gcd}\, (n_1,n_2)$ is large enough. The right panel depicts the flat direction in the fundamental domain for which the axion periodicity is manifest.
}
\label{fig1}
\end{figure}

To discuss the KNP mechanism, it is convenient to consider the limit 
$\Lambda_2=0$, in which the
axion potential is given by
\bea
V(\phi_i) =
\Lambda_1^4\left[1-\cos\left(\frac{n_1\phi_1}{f_1}+\frac{n_2\phi_2}{f_2}\right)\right].
\eea
Obviously this axion potential has a one-dimensional periodic  flat direction in the two-dimensional fundamental domain of the axion fields spanned by $\phi_i =[0, 2\pi f_i]$,
\bea
\phi_{\rm flat} \,\propto\, \frac{n_2\phi_1}{f_2} -\frac{n_1\phi_2}{f_1},\eea
 which can be identified as
the inflaton direction.
One easily finds that the length of this periodic flat direction is given by
\bea
\ell_{\rm flat} = \frac{2\pi \sqrt{ n_1^2 f_2^2 + n_2^2 f_1^2}}{{\rm gcd} \,(n_1, n_2)},
\label{flat.length}
\eea 
where ${\rm gcd}\, (n_1,n_2)$ denotes the greatest common divisor of $n_1$ and $n_2$.
This shows that a super-Planckian flat direction with  $\ell_{\rm flat} \gg   M_{Pl}\gg f_i$ can be developed  on the two-dimensional sub-Planckian domain if 
\bea
\frac{n_1}{{\rm gcd}\, (n_1, n_2)} \, \mbox{ or } \, \frac{n_2}{{\rm gcd}\, (n_1, n_2)}
\,\gg \, \frac{M_{Pl}}{f_i}\, \gg \, 1.\eea 
In Fig. \ref{fig1}, we depict the flat direction 
in the fundamental domain of axion fields, which has a length given by (\ref{flat.length}).
 Since the axionic inflaton of natural inflation rolls down along this periodic flat direction, its effective decay constant  is bounded as
 $$
 f_{\rm eff} \, \leq\, \frac{\ell_{\rm flat}}{2\pi},
 $$
which means that at least one of $n_i$ should be as large as
${\rm gcd}\,(n_1,n_2)f_{\rm eff}/f_i$.

Turning on the second axion potential
\bea
\Delta V =\Lambda_2^4\left[1-\cos\left(\frac{m_1\phi_1}{f_1}+\frac{m_2\phi_2}{f_2}\right)\right],\eea
a nontrivial potential is developed along the periodic flat direction having a length (\ref{flat.length}).
Even when $\ell_{\rm flat} \gg M_{Pl}$, 
%  flat direction (\ref{flat.length}) has a super-Planckian length, 
natural inflation is not guaranteed as the inflaton potential induced by $\Delta V$ generically has multiple
modulations along the flat direction. We find that the number of modulations  over the full range of the periodic flat direction 
is given by
\bea
N_{\rm mod}\, \,=\, \frac{|n_1m_2-n_2m_1|}{{\rm gcd}\, (n_1, n_2)},\eea
which results in the effective decay constant 
\bea
f_{\rm eff}\,=\, \frac{\ell_{\rm flat}}{2\pi N_{\rm mod}}\,=\, \frac{\sqrt{ n_1^2 f_2^2 + n_2^2 f_1^2}}{|n_1m_2-n_2m_1|}.\eea

It is straightforward to derive the above form of the effective axion decay constant \cite{Kim:2004rp}.
Taking the rotated axion field basis:
\bea
\psi =\frac{n_1 f_2 \phi_1 + n_2 f_1 \phi_2}{\sqrt{n_1^2 f_2^2 + n_2^2 f_1^2}}, \quad
\phi_{\rm flat}=\frac{n_2 f_1 \phi_1 - n_1 f_2 \phi_2}{\sqrt{n_1^2 f_2^2 + n_2^2 f_1^2}},\eea
%the degeneracy condition implies, the flatness of the inflaton direction crucially depends on the %anomaly coefficients. In order to see that, let us assume a hierarchy between amplitudes 
%$\Lambda_1^4\gg\Lambda_2^4$ for simplicity. Although it is unnecessary, the hierarchy assumption
% considerably simplifies analytic expressions for the inflaton field. Under the hierarchy of %amplitudes, an approximate heavy mass eigenstate becomes $\psi\sim(n_1\phi_1/f_1+n_2\phi_2/f_2)$ while %the orthogonal direction to $\psi$ direction becomes an approximate light mass eigenstate $\chi\sim(-%n_2\phi_1/f_2+n_1\phi_2/f_1)$ which corresponds to the inflaton direction. Changing field basis from $%(\phi_{1},\phi_{2})$ to $(\psi,\chi)$, 
the axion potential (\ref{potential.2axion}) can be written as
%we shall write  
\begin{eqnarray}
	V(\psi,\phi_{\rm flat})=\Lambda_1^4\left[1-\cos\left(\frac{\psi}{f_\psi}\right)\right]+\Lambda_2^4\left[1-\cos\left(\frac{\psi}{f'_\psi}+\frac{\phi_{\rm flat}}{f_{\rm eff}}\right)\right],
	\label{potential.2axion.changed}
\end{eqnarray}
where 
\begin{eqnarray}
	f_\psi&=&\frac{f_1f_2}{\sqrt{n_1^2f_2^2+n_2^2f_1^2}}, \nonumber \\
	f'_\psi&=&\frac{f_1f_2\sqrt{n_1^2f_2^2+n_2^2f_1^2}}{n_1m_1f_2^2+n_2m_2f_1^2}, \nonumber \\
	f_{\rm eff}&=&\frac{\sqrt{n_1^2f_2^2+n_2^2f_1^2}}{|n_1m_2-n_2m_1|}.
	\label{effective.decay.const}
\end{eqnarray}
Taking the limit $\Lambda_1^4 \gg \Lambda_2^4$, the heavy field component $\psi$
can be integrated out, yielding the effective potential of the light inflaton direction $\phi_{\rm flat}$ as
%Since $\psi$ field is heavier than $\chi$ field, it rapidly roll down to its minimum point,
% $\langle\psi\rangle=2\pi n f_\psi$. After $\psi$ field gets its vacuum expectation value, the %effective potential for $\chi$ field is generated
\begin{eqnarray}
	V_{\textrm{eff}}(\phi_{\rm flat})=\Lambda_2^4\left[ 1-\cos\left( \frac{\phi_{\rm flat}}{f_{\rm eff}} \right) \right].
	\label{effectchi}
\end{eqnarray}
%up to additional phase factor $\delta$, which is irrelavant for inflationary dynamics. Although its %potential form is same as single field natural inflation, its decay constant $f_\chi$ is not the %fundamental scale emerged from $U(1)_{PQ}$ symmetry breaking: it is a effectively generated parameter %as a function of anomaly coefficients and the decay constant $f_{1,2}$ so that it could exceed even %the Planck scale $M_{\textrm{pl}}$. For the string axions of comparable decay constant

From the above expression of $f_{\rm eff}$, it is clear that $f_{\rm eff}\gg f_i$ is {\it not} a generic feature of the model, but requires a special alignment  of the anomaly coefficients
$n_i, m_i$. Note that in the absence of any alignment, generically \bea
f_{\rm eff} \sim \frac{f_i}{n},\eea
where $n$ represents the typical size of the anomaly coefficients $n_i,m_i$.
Obviously there can be many different choices (or alignments) of the anomaly coefficients yielding $f_{\rm eff}/f_i \gg 1$.
A simple possibility is that one or both of $n_i$ are large, while the denominator $(n_1m_2-n_2m_1)$
is tuned to be ${\cal O}(1)$. As a specific example, KNP considered the case that 
$n_2\gg 1$ and the other three anomaly coefficients are given by  
$m_1=n_1=1$ and $m_2= n_2+{\cal O}(1)$~\cite{Kim:2004rp}. Of course, one can consider different examples as we do below, for instance
$n_2\gg 1$ with $n_1=m_2=1$, $m_1=0$.

To understand the geometric meaning of the required alignment, one can introduce an alignment angle $\delta\theta$ as
\bea
\sin \delta\theta \,\equiv \,  \frac{|n_1m_2-m_1n_2|}{\sqrt{(n_1^2+n_2^2)(m_1^2+m_2^2)}},\eea
and rewrite
 the effective decay constant as
\bea
f_{\rm eff} \,=\, \frac{1}{\sin\delta\theta}\left(\frac{n_1^2f_2^2+n_2^2 f_1^2}{(n_1^2+n_2^2)(m_1^2+m_2^2)}\right)^{1/2}.\eea
Note that $\delta\theta$ corresponds to the angle  between the heavy field direction $\psi$ and 
the other field direction $\phi$ of the second axion potential $\Delta V$: 
\bea
\psi \,\propto\, \frac{n_1\phi_1}{f_1}+\frac{n_2\phi_2}{f_2}, \quad \phi\,\propto\, \frac{m_1\phi_1}{f_1}+\frac{m_2\phi_2}{f_2}.\eea
%and this explains why small $\epsilon_d$ yields a large $f_{\rm eff}/f_i$.
For the case of two axions, one of the economic  ways to get $f_{\rm eff}/f_i\gg 1$ is  to have
\bea
n_1 \,\sim\, n_2 \,=\, {\cal O}({f_{\rm eff}}/{f_{i}}), \quad m_1\,\sim\, m_2 \,=\, {\cal O}(1),
\label{region1} 
\eea
for which
\bea
\delta\theta  \,=\, {\cal O}(f_i/f_{\rm eff}).\eea
This implies that the probability for achieving $f_{\rm eff}\gg f_i$
under {a random choice of the anomaly coefficients}, but within the specific region of  (\ref{region1}),
is given by
\bea
P(f_{\rm eff}/f_i) ={\cal O}(f_i/f_{\rm eff}).
\eea
On the other hand, if one extends the random choice to the generic region where all the anomaly coefficients can be of the order of $f_{\rm eff}/f_i$, one needs an alignment angle  $\delta\theta={\cal O}(f_i^2/f_{\rm eff}^2)$, and then the probability for achieving $f_{\rm eff}\gg f_i$ is reduced as
\bea
P(f_{\rm eff}/f_i) \,=\, {\cal O}(f_i^2/f^2_{\rm eff}).\eea

A particularly interesting choice 
\cite{Tye:2014tja,Ben-Dayan:2014zsa} 
 of the anomaly coefficients realizing  the KNP alignment 
is
\bea
n_1=m_2=1, \quad m_1=0, \quad |n_2|\gg 1,\eea
for which the light inflaton direction is identified as
\bea
\phi_{\rm flat} \,=\, \frac{n_2 f_1 \phi_1-f_2\phi_2}{f_{\rm eff}} \,\simeq\, \phi_1
\eea
with the effective decay constant 
\bea
f_{\rm eff} \,=\, \sqrt{n_2^2 f_1^2 + f_2^2} \,\simeq\, |n_2| f_1\eea
where we assumed $f_1\sim f_2$. In this case, 
we have
\bea
e^{i\phi_1/ f_1} \,=\, e^{-in_2{\phi_2}/{f_2}}\eea
along the inflaton direction. Then the enhanced effective axion decay constant   can be considered as
a consequence of the $Z_{n_2}$ monodromy structure of $\phi_1$, which is induced  by the mixing with the heavy axion component
$\phi_2$.

\section{Generalization to $N >$ 2 axions}\label{sec:3}

It is in fact straightforward to generalize the KNP mechanism to the case with $N >2$ axions.\footnote{
A generalization to the case with $N>2$ axions has been discussed in~\cite{Chatzistavrakidis:2012bb} to accommodate 
the intermediate scale QCD axion as well as a quintessence axion with Planck scale decay constant in the model.}
In the presence of $N$ axions, again in the periodic axion field basis
\bea
\phi_i \,\equiv \, \phi_i + 2\pi f_i \quad (i=1,2,\ldots, N),\eea
  the axion potential takes the form 
\begin{eqnarray}
	V=\sum_{i=1}^N \Lambda_i^4\left[1-\cos\left(\sum_{j=1}^N \frac{n_{ij}\phi_j}{f_j}\right)\right],
	\label{potential.Naxion}
\end{eqnarray}
where $\vec{n_i}=(n_{i1}, n_{i2}, .., n_{iN})$ are linearly independent integer-valued anomaly coefficients, and $f_i\ll M_{Pl}$ are the fundamental decay constants
which are presumed to be comparable to each other.  
To proceed, let us take the limit 
$$\Lambda_N=0,$$ 
for which the one-dimensional periodic flat direction is parametrized as
\begin{eqnarray}
	\phi_{\rm flat} &\propto & \sum_i X_i \phi_i 
	\nonumber \\
&\equiv &	\textrm{Det}\left(
	\begin{tabular}{cccc}
	${\phi_1}$&${\phi_2}$&$\cdots$&${\phi_N}$\\
	$\frac{n_{11}}{f_1}$&$\frac{n_{12}}{f_2}$&$\cdots$&$\frac{n_{1N}}{f_N}$\\
	$\vdots$&$\vdots$&&$\vdots$\\
	$\frac{n_{N-1,1}}{f_1}$&$\frac{n_{N-1,2}}{f_2}$&$\cdots$&$\frac{n_{N-1,N}}{f_N}$
	\end{tabular}
	\right),
	\label{inflaton.direction}
\end{eqnarray}
where 
\bea
X_i \,=\, \frac{C_if_i}{\prod_j f_j}\eea
% $\vec{C}=\left(C_1,C_2,\cdots,C_N\right)$ is
for
\begin{eqnarray}
	C_i=\left(-1\right)^{i+1}\textrm{Det}\left(\begin{array}{ccccccc}
		n_{11} & n_{12} & \ldots & n_{1,(i-1)} & n_{1,(i+1)} & \ldots & n_{1N} \\
		n_{21} & n_{22} & \ldots & n_{2,(i-1)} & n_{2,(i+1)} & \ldots & n_{2N} \\
		\vdots & \vdots & & \vdots & \vdots & & \vdots \\
		n_{N-1,1} & n_{N-1,2} & \ldots & n_{N-1,(i-1)} & n_{N-1,(i+1)} & \ldots & n_{N-1,N}
		\end{array}\right).
		\label{numerator}
\end{eqnarray}

The length of this periodic flat direction is determined by the minimal discrete shift
$\Delta \phi_i$ along the flat direction, under which the axion field configuration comes back to
the original  configuration.
One then finds
 \bea
\Delta \phi_i \,=\, \frac{2\pi C_if_i}{{\rm gcd}\, (C_1,C_2,\cdots, C_N)},\eea
 yielding the length of the flat direction:
\begin{eqnarray}
\ell_{\rm flat }=\frac{2\pi \sqrt{\sum_{i=1}^N C_i^2f_i^2}}{{\rm gcd}\,\left(C_1,C_2,\cdots, C_N\right)},
\label{scale.inflaton}
\end{eqnarray}
where ${\rm gcd}\left(C_1,C_2,\cdots, C_N\right)$ denotes the greatest common divisor of all $C_i$. 

For generic anomaly coefficients $n_{ij}$, the magnitude  of $C_i$
is quite sensitive to the number of axions, as well as to the typical size of  $n_{ij}$.  Here we are interested in the limit $N\gg 1$ with 
\bea
|n_{ij}|\leq n_{\rm max}={\cal O}({\rm few}). \label{max}
\eea
 To proceed, we can regard $n_{ij}$ as a random integer-valued variable with flat
		probability distribution:
		$$ P(n_{ij})=\frac{1}{2n_{\rm
		max}+1}.$$  We then have 
		\bea
		\langle n_{ij} \rangle &=& \sum_{n_{ij}=-n_{\rm max}}^{n_{\rm max}} n_{ij}P(n_{ij}) \,=\, 0,\nonumber \\
		\langle n_{ij}^2 \rangle &=& \sum_{n_{ij}=-n_{\rm max}}^{n_{\rm max}}
		n_{ij}^2P(n_{ij}) \,= \,\frac{1}{3}n_{\rm max}(1+n_{\rm
		max})\,\equiv\, n^2,\nonumber \eea 
where 
$$
n\,=\, \sqrt{n_{\rm max}(1+n_{\rm max})/3}
$$
denotes the typical size of the anomaly coefficients in the range (\ref{max}). 		
		One can similarly compute the expectation value of 
		$C_i^{\,2}$. For 
		$C_N = \sum_{\sigma}{\rm sgn}(\sigma)
		n_{1\sigma(1)}\cdots n_{(N-1)\sigma(N-1)}$,
		where the summation is over all possible permutations
		of $\{1,2,\ldots,N-1\}$, one easily finds 
		$$\langle C_N^2\rangle = \left \langle\sum_{\sigma}
	(n_{1\sigma(1)}\cdots n_{(N-1)\sigma(N-1)})^2\right\rangle = (N-1)!\, n^{2(N-1)},$$
and therefore
		$$\left\langle \sum_{i=1}^N C_i^2 \right\rangle = N!\cdot n^{2(N-1)}.$$

As implied by this expectation value, in most cases we 
have\footnote{We have in fact an upper bound $\sqrt{\sum_i C_i^2} < N^{N/2}n_{\rm max}^{N-1}$.
We found through a numerical analysis that $\sqrt{\sum_i C_i^2} \gtrsim 0.2 \sqrt{N!}\,n^{N-1}$ in most cases of our interest.  It is also known that a probability for ${\rm gcd}\,(C_i)=1$ under a random choice of $C_i$ within the range
$|C_i|\leq Q$ is given by $P({\rm gcd}(C_i) =1)=1/\zeta(N) +{\cal O}(1/Q),$ which is close to the unity in the limit $N\gg 1$ and $Q\gg 1$. Although in our case 
$C_i$ is not a randomly chosen integer, but a determinant of the randomly chosen anomaly coefficients $n_{ij}$, we confirmed again through a numerical analysis that  ${\rm gcd}(C_i) =1$ in most cases of our interest.} 
\bea 
\left(\sum_{i=1}^N C_i^2\right)^{1/2} \,\sim \, \sqrt{N!}\, n^{N-1}, \quad
{\rm gcd}\,(C_1,C_2,\ldots,C_N) \,=\, {\cal O}(1),\eea
and therefore a flat direction  enhanced as
 \bea
 \frac{\ell_{\rm flat}}{2\pi f_i}  \,\sim \, \sqrt{N!}n^{N-1},\eea
 where the sub-Planckian axion decays constants $f_i$ are assumed to be comparable to each other.
 Here we are interested in the case that the sub-Planckian axions $\phi_i$ originate from 
 higher-dimensional tensor gauge fields in compactified string theory,  in which case all $f_i$ are comparable to $M_{st}/8\pi^2$ for the string scale $M_{st}$
 \cite{Choi:1985je}.
  Note that the flat direction length is exponentially enhanced in the limit $N\gg 1$ when the typical anomaly coefficient
 $n>1$.    
%which means that 
%an exponentially long flat  direction is a generic feature of the multiple axion models in the limit $N\gg 1$, even when the %anomaly coefficients are limited to be within the order of unity.
As a result,   an exponentially long flat direction can be obtained with
the anomaly coefficients $|n_{ij}| \leq n_{\rm max} ={\cal O}({\rm few})$, with
a moderately large number of axions
\bea
N \ln N  \sim 2\ln( \ell_{\rm flat}/2\pi f_i).\eea
This can be understood by noting that the flat direction generically winds each of the additional axion dimensions by about $n$ times, which would explain the factor $n^{N-1}$, and there are also combinatoric degrees of freedom for the windings of the $N$-dimensional torus of axion fields, which 
would explain the factor $\sqrt{N!}$.

Introducing the last part of the axion potential
$$
\Delta V \,=\, \Lambda_N^4 \left[ 1-\cos\left(\sum_j \frac{n_{Nj}\phi_j}{f_j}\right)\right],
$$
a periodic potential is developed along the flat direction. 
Again  super-Planckian flat direction does not guarantee a super-Planckian effective decay constant. Instead we have
\bea
f_{\rm eff}\,=\, \frac{\ell_{\rm flat}}{2\pi N_{\rm mod}}, \eea
where $N_{\rm mod}$ is the number of modulations of the axion potential over the full period of the flat direction.
Taking the limit $\Lambda_i\gg \Lambda_N$ ($i=1,2,\ldots, N-1$) to integrate out the $(N-1)$ heavy axions,  we find that the effective potential of the flat direction is given by
%  decay constant of the flat direction  $\phi_{\rm flat}$ as
\bea
V_{\rm eff}(\phi_{\rm flat})\,=\, \Lambda_N^4 \left[1-\cos\left(\frac{\phi_{\rm flat}}{f_{\rm eff}}\right)\right],\eea
where 
\begin{eqnarray}
f_{\rm eff}\,=\,\frac{\sqrt{\sum_{i=1}^{N}C_i^2f_i^2}}{|\textrm{Det}\, \mathbb{N}|},
\label{decay.constant.inflaton}
\end{eqnarray}
for
\begin{eqnarray}
{\rm Det}\,\mathbb{N}= {\rm Det}\left(
	\begin{tabular}{cccc}
	$n_{N1}$&$n_{N2}$&$\cdots$&$n_{NN}$\\
	$n_{11}$&$n_{12}$&$\cdots$&$n_{1N}$\\
	$\vdots$&$\vdots$&&$\vdots$\\
	$n_{(N-1)1}$&$n_{(N-1)2}$&$\cdots$&$n_{(N-1)N}$
	\end{tabular}
	\right)=\sum_i C_in_{Ni}.
\end{eqnarray}
%where $\vec{C}$ is defined by \eqref{numerator}.
This tells that the number of modulations over the full range of the flat direction
is 
$$
N_{\rm mod}\,=\, \frac{\left|\sum_i C_in_{Ni}\right|}{{\rm gcd}\, (C_1,C_2, \cdots, C_N)}.$$

To justify our approach to integrate out the $(N-1)$ heavy axions, let us  briefly examine the axion masses
in our scheme. For the axion potential (\ref{potential.Naxion}), the $N\times N$ axion mass-square matrix is given by  
		$m^2_{kl} = \sum_i \Lambda_i^4
		n_{ik}n_{il}/f_kf_l$, yielding 
		 \bea {\rm Det}\, m^2 =  ({\rm Det}\,\mathbb{N})^2\prod_i^N\frac{
		\Lambda_i^4}{ f_i^2}   .\eea
		In the presence of light flat direction, its mass square is given by 
		\bea m^2_{\rm flat}\,\simeq \, \frac{\Lambda_N^4}{f_{\rm eff}^2} \,=\,\frac{\Lambda_N^4 ({\rm Det}\,\mathbb{N})^2 }{\sum_j C_j^2 f_j^2}.\eea
		Then the mass-square  determinant can be written as
		\bea
		{\rm Det}\, m^2 &=& ({\rm Det}\,\mathbb{N})^2 \prod_i^N\frac{
		\Lambda_i^4}{ f_i^2}  \,=\, \frac{\Lambda_N^4}{f_{\rm
		eff}^2}\frac{f_{\rm eff}^2}{f_N^2}({\rm Det}\,\mathbb{N})^2
		\prod_{i}^{N-1}  \frac{\Lambda_i^4}{
		f_i^2} \nonumber \\ 
		&=&  m^2_{\rm flat} \frac{\sum C_j^2 f_j^2}{f_N^2}\prod_i^{N-1} \frac{\Lambda_i^4}{f_i^2}
		 \,\sim\,  m_{\rm flat}^2\prod_{i}^{N-1}  \frac{N n^2 \Lambda_i^4}{e
		f_i^2} \,\sim\, m_{\rm flat}^2\prod_i^{N-1} m^2_{\rm heavy}(i),
		\eea
		where $m_{\rm heavy}(i)$ is the $i$th  heavy axion mass, and we have used $\sum_j C_j^2 f_j^2/f_N^2 \sim N!n^{2(N-1)}\sim N^Nn^{2(N-1)}/e^N$ under the assumption that all $f_i$ are comparable to each other. 
		We then find  
		\bea
		\frac{m^2_{\rm heavy}(i)}{m^2_{\rm flat}}\, \sim\, \frac{Nn^2}{e} \frac{f_{\rm eff}^2}{f_i^2}\frac{\Lambda_i^4}{\Lambda_N^4},\eea 
		which shows that the heavy  axions are heavy enough compared to the flat direction if
		the anomaly coefficients are aligned to yield $f_{\rm eff}\gg f_i$,
		even when
		$\Lambda_N$ is comparable to
		$\Lambda_i$ ($i=1,..., N-1$).

As in the case with two axions,  it is clear that $f_{\rm eff}\gg f_i$ is not a generic feature, but requires a specific alignment of the anomaly coefficients. Yet, compared to the case of two axions, a notable difference is that
the mechanism does not require large anomaly coefficients of ${\cal O}(f_{\rm eff}/f_i)$, but
a moderately large number of axions $N\ln N\gtrsim 2\ln (f_{\rm eff}/f_i)$
together with the anomaly coefficients $n_{ij}={\cal O}(1)$. 
To quantify the required degree of alignment, let us introduce an alignment angle as in the case of two axions:
\bea
\sin \delta\theta \,\equiv \, \frac{|\vec{C}\cdot \vec{n}_N|}{|\vec{C}||\vec{n_N}|} \,=\, \frac{|{\rm Det}\,\mathbb{N}|}{|\vec{C}||\vec{n}_N|}.\eea
Then the effective decay constant reads as
\bea
f_{\rm eff} \,=\, \frac{1}{\sin\delta\theta}\frac{\sqrt{\sum_{i=1}^N C_i^2 f_i^2}}{|\vec{C}||\vec{n}_N|} \, \sim\, \frac{f_i}{|\vec{n}_N|\sin\delta\theta},\eea
implying that we need to align  $\delta\theta$ to be small as
\bea
\delta\theta \,=\, {\cal O}(f_i/f_{\rm eff}).\eea 
This also suggests that the probability for having $f_{\rm eff}\gg f_i$
under a random choice of the anomaly coefficients in the range
$|n_{ij}|\leq n_{\rm max}={\cal O}(1)$ is  given by
\bea
P(f_{\rm eff}/f_i) \,=\, {\cal O}(f_i/f_{\rm eff}).\eea
Note that  in the case of two axions we have a similar probability
only  when the random selection is limited to a specific (economic) region of the anomaly coefficients, for instance the region of (\ref{region1}), while $P(f_{\rm eff}/f_i)={\cal O}(f_i^2/f_{\rm eff}^2)$ if one extends the random selection to the general region where all the anomaly coefficients can be comparable to each other.  
%In the case with $N\gtrsim \ln(f_{\rm eff}/f_i)$, 

In view of that the fundamental axion decay constants in string theory are typically in the range $f_i\sim 10^{16}- 10^{17}$ GeV~\cite{Choi:1985je},
while a successful natural inflation compatible with the recent BICEP2 results~\cite{Ade:2014xna} requires
  $f_{\rm eff} \gtrsim 10M_{Pl}$, we are  particularly interested in the {\it minimal}
  number of axions  which can yield
\bea
f_{\rm eff}/f_i \,=\, {\cal O}(10^2-10^3) \mbox{ \, for \,}
|n_{ij}|\leq n_{\rm max} \,\, (n_{\rm max}=1, 2, 3).
\eea
According to our discussion above, the corresponding range of $N$ is roughly given by
\bea
N \ln N \,\gtrsim\, 2\ln (f_{\rm eff}/f_i).\eea
We have performed a numerical analysis to evaluate $P(f_{\rm eff}/f_i)$ for the three different values of $f_{\rm eff}/f_i$:
$$f_{\rm eff}/f_i \, =10^2,\,\, 5\times 10^2, \,\, 10^3,
$$ when 
$$ N=  8-15 \,\,  (n_{\rm max}=1), \quad N=7-14 \,\,(n_{\rm max}=2), \quad
N=5-12\,\, (n_{\rm max}=3).
$$ The results are depicted in Table~\ref{numerical.prob}, which confirms that
the probability for the necessary alignment to be achieved under a random choice of the anomaly coefficients is indeed
of the order of $f_i/f_{\rm eff}$.

\begin{table}
	\begin{minipage}{5cm}
	\setlength{\tabcolsep}{9pt}
        \begin{tabular}{|c|c|c|c|}
                \hline
		& \multicolumn{3}{c|}{$f_\textrm{eff}/f_i$} \\ \cline{2-4}
		\raisebox{2ex}{$N$} & 100 & 500 & 1000 \\ \hline \hline
                8 & 0.009 & 0. & 0. \\
                9 & 0.064 & 0. & 0. \\
                10 & 0.258 & 0.020 & 0. \\
                11 & 0.487 & 0.105 & 0.01 \\
                12 & 0.707 & 0.275 & 0.12 \\
                13 & 0.797 & 0.500 & 0.41 \\
                14 & 0.938 & 0.800 & 0.56 \\
                15 & 0.855 & 0.850 & 0.77 \\
                \hline
        \end{tabular}
	\end{minipage}
	\hspace{0.25cm}
	\begin{minipage}{5cm}
	\setlength{\tabcolsep}{9pt}
	\begin{tabular}{|c|c|c|c|}
		\hline
			& \multicolumn{3}{c|}{$f_\textrm{eff}/f_i$}  \\ \cline{2-4}
		\raisebox{2ex}{$N$} & 100 & 500 & 1000 \\ \hline \hline
		7 & 0.216 & 0.030 & 0. \\
		8 & 0.411 & 0.210 & 0.05 \\
		9 & 0.466 & 0.530 & 0.26 \\
		10 & 0.542 & 0.445 & 0.43 \\
		11 & 0.512 & 0.470 & 0.70 \\
		12 & 0.519 & 0.660 & 0.64 \\
		13 & 0.585 & 0.585 & 0.47 \\
		14 & 0.530 & 0.490 & 0.43 \\
		\hline
	\end{tabular}
	\end{minipage}
\hspace{0.25cm}
	\begin{minipage}{5cm}
	\setlength{\tabcolsep}{9pt}
	\begin{tabular}{|c|c|c|c|}
		\hline
		& \multicolumn{3}{c|}{$f_\textrm{eff}/f_i$} \\ \cline{2-4}
		\raisebox{2ex}{$N$} & 100 & 500 & 1000 \\ \hline \hline
		5 & 0.060 & 0. & 0. \\
		6 & 0.202 & 0.060 & 0. \\
		7 & 0.322 & 0.230 & 0.09 \\
		8 & 0.373 & 0.225 & 0.31 \\
		9 & 0.383 & 0.370 & 0.40 \\
        10 & 0.393 & 0.390 & 0.30 \\
        11 & 0.408 & 0.370 & 0.41 \\
        12 & 0.404 & 0.355 & 0.39 \\
		\hline
	\end{tabular}	
	\end{minipage}			

	\caption{The probability to $f_i/f_{\rm eff}$ ratio, $R=P/(f_i/f_{\rm eff})$, for the necessary alignment under  $10^5$ random choices of the anomaly coefficients. We have considered  $N=8-15$ for
	$n_{\rm max}=1$,  $N=7-14$ for $n_{\rm max}=2$, and $N=5-12$ for $n_{\rm max}=3$.}
\label{numerical.prob}
\end{table}

Before closing this section, let us present a couple of explicit models which achieve an exponentially enhanced effective axion decay constant within the framework discussed above.
Our first model is  
\begin{eqnarray}
	V &=& \Lambda_1^4\left[ 1-\cos\left( \frac{\phi_1}{f_1}+\frac{n_2 \phi_2}{f_2}\right) \right]+\Lambda_2^4\left[ 1-\cos\left( \frac{\phi_2}{f_2}+\frac{n_3\phi_3}{f_3}\right) \right]+\cdots \\ \nonumber
	&+&\Lambda_{N-1}^4\left[ 1-\cos\left( \frac{\phi_{N-1}}{f_{N-1}}+\frac{n_{N}\phi_N}{f_N} \right) \right] +\Lambda_N^4\left[ 1-\cos\left(\frac{\phi_N}{f_N}\right) \right],
	\label{Potential.Ex1}
\end{eqnarray}
which is designed to realize a multiple axion monodromy  along the
inflaton direction.  
The  anomaly coefficient matrix of the model takes the form
\begin{eqnarray}
	\mathbb{N} &=&  \left(n_{ij}\right) \,=\, \left(
	\begin{array}{cccccc}
		1 & n_2 &   & 	    &   &   \\
		  & 1 & n_3 & 	    &   &   \\
		  &   &  1 & \ddots  &   &   \\
		  &   &   &   \ddots      &  & n_N\\
		  &   &   &         &   & 1 \\
	\end{array}
	\right),
	\label{Ex1}
\end{eqnarray}
for which
\begin{eqnarray}
	\textrm{Det}\,\mathbb{N} =1, \quad
	\left|C_i\right| = \prod_{j=i+1}^{N}n_j.
	\label{Ci}
\end{eqnarray}
The resulting effective axion decay constant is given by
\begin{eqnarray}
	f_{\rm eff} \,= \, \frac{\sqrt{\sum_{i=1}^N f_i^2 C_i^2}}{\textrm{Det}\,\mathbb{N}} 
	\,=\, \left( \sum_{i=1}^N \prod_{j=i+1}^{N}n_j^2 f_i^2 \right)^{1/2}\,\sim\, n_2n_3\cdots n_N f_1
	\label{Eff.Decay}
\end{eqnarray}
if $n_i > 1$ and $f_1\sim f_2 \sim \cdots \sim f_N$. 
In the limit $\Lambda_N \ll \Lambda_i$ ($i=1,\ldots,N-1$),
the inflaton direction is determined to be 
\begin{eqnarray}
	\phi_{\rm flat}\, \propto \,  \sum_i C_i\phi_i 
	\,=\, \left(\prod_{i=2}^{N}n_{i}\right)\phi_1 - \left(\prod_{i=3}^{N}n_{i}\right)\phi_2 +\cdots - n_{N}\phi_{N-1}+ \phi_N
	\label{Eff.Dirc}
\end{eqnarray}

\begin{figure}[t]
	\centering
	\includegraphics{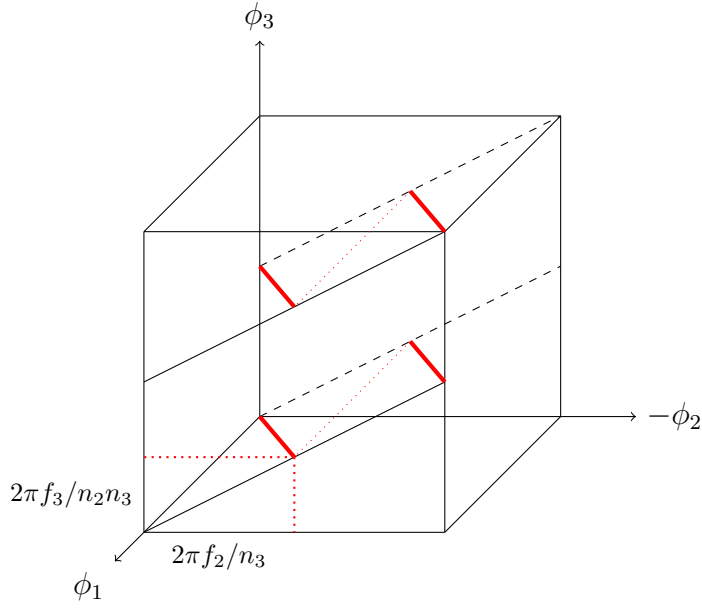}
	\caption{Multiple monodromy structure for the three-axion model with  $n_2=n_3=2$. The solid red line represents the inflaton direction
	in the fundamental domain of three axions. Note that $\Delta \phi_2=2\pi f_2$ along the inflaton direction requires  $\Delta \phi_1=2\pi n_2f_1$, and
	  $\Delta \phi_3=2\pi f_3$ requires $\Delta \phi_2=2\pi n_3 f_2$. As a result, $\Delta \phi_3=2\pi f_3$ along the inflaton direction yields $\Delta \phi_1=2\pi n_2n_3 f_1$.}
	\label{fig:Dirc.3D}
\end{figure}

This  model can be considered as 
a generalization of the
two axion models of \cite{Tye:2014tja,Ben-Dayan:2014zsa}, and  realizes a multiple axion monodromy  $\prod_{i=2}^N Z_{n_i}$  along the inflaton direction. As a consequence,
in order for the $N$th axion  $\phi_N$ to travel one period  along the  inflaton direction, i.e. $\Delta \phi_N = 2\pi f_N$, the other axions $\phi_i$ ($i=1,2,\ldots, N-1$) should experience a multiple winding
as
\begin{eqnarray}
\frac{\Delta \phi_i}{2\pi f_i}  \,=\, \prod_{j\geq i+1}^N n_j.\end{eqnarray} 
In Fig.~\ref{fig:Dirc.3D}, we depict such multiple monodromy structure for the case of $N=3$ and $n_2=n_3=2$.

Our second model is 
\begin{eqnarray}
	V &=& \Lambda_1^4\left[ 1-\cos\left( \frac{\phi_1}{f_1}+\frac{\phi_2}{f_2}-\frac{\phi_3}{f_3}+\cdots+(-)^{N}\frac{\phi_N}{f_N}\right) \right]
	\nonumber \\ \nonumber
	&+&\Lambda_2^4\left[ 1-\cos\left( \frac{\phi_2}{f_2}+\frac{\phi_3}{f_3}-\frac{\phi_4}{f_4}+\cdots+(-)^{N-1}\frac{\phi_N}{f_N}\right) \right] +\cdots\\ \nonumber
	& + &\, \Lambda_{N-2}^4\left[ 1-\cos\left( \frac{\phi_{N-2}}{f_{N-2}}+\frac{\phi_{N-1}}{f_{N-1}} -\frac{\phi_N}{f_N}\right) \right]
\nonumber \\	
&+&	\Lambda_{N-1}^4\left[ 1-\cos\left( \frac{\phi_{N-1}}{f_{N-1}}+\frac{\phi_N}{f_N} \right) \right] +\Lambda_N^4\left[ 1-\cos\left(\frac{\phi_N}{f_N}\right) \right],
	\label{Potential.Ex2}
\end{eqnarray}
which is designed to achieve an exponentially enhanced effective decay  constant even when all the integer-valued  anomaly
coefficients are restricted as $|n_{ij}|\leq 1$.
In this case, we have
\begin{eqnarray}
\textrm{Det}\,\mathbb{N} =1,  \quad  C^2_{i}=4^{(N-1-i)}\quad (i=1,\,2,\,\cdots,\,N-1), \quad
	C_N^2=1.
	\label{Sub.Determinant}
\end{eqnarray}
Then, assuming that all fundamental decay constants are comparable to each other, the effective decay constant is exponentially enhanced as 
\begin{eqnarray}
	f_{\rm eff}\,\sim\, \sqrt{\frac{1}{3}(4^{N-1}+2)}\,f_i \,\sim\,  \frac{2^{N-1}}{\sqrt{3}}f_i
	\label{Eff.Decay.Ex2}
\end{eqnarray}
although we have  $|n_{ij}|\leq 1$. 

\section{Conclusion}\label{sec:4}

Natural inflation provides an attractive framework for large field chaotic inflation which can explain the recent detection of primordial gravitational waves by BICEP2. The KNP alignment mechanism offers an interesting scheme to get a super-Planckian effective axion decay constant $f_{\rm eff}\gg M_{Pl}$, which is a necessary component of natural inflation, starting from sub-Planckian fundamental axion decay constants $f_i\ll M_{Pl}$ of multiple axions. In this paper, we extended the original KNP model with two axions
to models with $N>2$ axions. Compared to the original KNP model, a notable difference 
is that large anomaly coefficients of ${\cal O}(f_{\rm eff}/f_i)$ are not required anymore if
the number of axions is moderately large as $N\ln N\gtrsim 2\ln (f_{\rm eff}/f_i)$. 
With such $N$, 
the effective decay constant can be exponentially enhanced as $f_{\rm eff}/f_i \sim \sqrt{N!}\,n^{N-1}$ for $n$ denoting the typical size of the integer-valued anomaly coefficients, and
the probability for the necessary alignment to be achieved under a random choice of the anomaly coefficients is 
of the order of $f_i/f_{\rm eff}$.

The structure of our setup is rich enough to realize a variety of different possibilities.  For instance,  it
can realize a multiple axion monodromy $\prod_{i=2}^N Z_{n_i}$ yielding $f_{\rm eff}/f_i\sim \prod_{i=2}^N n_i$. The setup can also give rise to 
an exponentially enhanced effective axion decay constant as
 $f_{\rm eff}/f_i \sim 2^{N-1}$, even when all the integer-valued anomaly coefficients are restricted as  $|n_{ij}|\leq 1$.

\section*{Acknowledgements}We thank Chang Sub Shin, Kwang Sik Jeong and Jinn Ouk Gong for helpful discussions.


\begin{thebibliography}{99}
%\cite{Ade:2014xna}
\bibitem{Ade:2014xna} 
  P.~A.~R.~Ade {\it et al.}  [BICEP2 Collaboration],
  %``Detection of B-Mode Polarization at Degree Angular Scales by BICEP2,''
  Phys.\ Rev.\ Lett.\  {\bf 112}, 241101 (2014)
  [arXiv:1403.3985 [astro-ph.CO]].
  %%CITATION = ARXIV:1403.3985;%%
  %596 citations counted in INSPIRE as of 10 Aug 2014

%\cite{Lyth:1996im}
\bibitem{Lyth:1996im} 
  D.~H.~Lyth,
  %``What would we learn by detecting a gravitational wave signal in the cosmic microwave background anisotropy?,''
  Phys.\ Rev.\ Lett.\  {\bf 78}, 1861 (1997)
  [hep-ph/9606387].
  %%CITATION = HEP-PH/9606387;%%
  %303 citations counted in INSPIRE as of 22 Apr 2014

%\cite{Freese:1990rb}
\bibitem{Freese:1990rb} 
  K.~Freese, J.~A.~Frieman and A.~V.~Olinto,
  %``Natural inflation with pseudo - Nambu-Goldstone bosons,''
  Phys.\ Rev.\ Lett.\  {\bf 65}, 3233 (1990);
  F.~C.~Adams, J.~R.~Bond, K.~Freese, J.~A.~Frieman and A.~V.~Olinto,
  %``Natural inflation: Particle physics models, power law spectra for large scale structure, and constraints from COBE,''
  Phys.\ Rev.\ D {\bf 47}, 426 (1993)
  [hep-ph/9207245].
  %%CITATION = HEP-PH/9207245;%%
  %323 citations counted in INSPIRE as of 22 Apr 2014

%\cite{Freese:2014nla}
\bibitem{Freese:2014nla} 
  K.~Freese and W.~H.~Kinney,
  %``Natural Inflation: Consistency with Cosmic Microwave Background Observations of Planck and BICEP2,''
  arXiv:1403.5277 [astro-ph.CO].
  %%CITATION = ARXIV:1403.5277;%%
  %13 citations counted in INSPIRE as of 22 Apr 2014

%\cite{Choi:1985je}
\bibitem{Choi:1985je} 
  K.~Choi and J.~E.~Kim,
  %``Harmful Axions in Superstring Models,''
  Phys.\ Lett.\ B {\bf 154}, 393 (1985)
  [Erratum-ibid.\  {\bf 156B}, 452 (1985)];
   P.~Svrcek and E.~Witten,
  %``Axions In String Theory,''
  JHEP {\bf 0606}, 051 (2006)
  [hep-th/0605206];
  T.~Banks, M.~Dine, P.~J.~Fox and E.~Gorbatov,
  %``On the possibility of large axion decay constants,''
  JCAP {\bf 0306}, 001 (2003)
  [hep-th/0303252].
  %%CITATION = HEP-TH/0303252;%%
  %101 citations counted in INSPIRE as of 24 Apr 2014

%\cite{Kim:2004rp}
\bibitem{Kim:2004rp} 
  J.~E.~Kim, H.~P.~Nilles and M.~Peloso,
  %``Completing natural inflation,''
  JCAP {\bf 0501}, 005 (2005)
  [hep-ph/0409138].
  %%CITATION = HEP-PH/0409138;%%
  %81 citations counted in INSPIRE as of 22 Apr 2014

%\cite{ArkaniHamed:2003wu}
\bibitem{ArkaniHamed:2003wu} 
  N.~Arkani-Hamed, H.~-C.~Cheng, P.~Creminelli and L.~Randall,
  %``Extra natural inflation,''
  Phys.\ Rev.\ Lett.\  {\bf 90}, 221302 (2003)
  [hep-th/0301218].
  %%CITATION = HEP-TH/0301218;%%
  %96 citations counted in INSPIRE as of 22 Apr 2014


%\cite{Germani:2010hd}
\bibitem{Germani:2010hd} 
  C.~Germani and A.~Kehagias,
  %``UV-Protected Inflation,''
  Phys.\ Rev.\ Lett.\  {\bf 106}, 161302 (2011)
  [arXiv:1012.0853 [hep-ph]];
  %%CITATION = ARXIV:1012.0853;%%
  %53 citations counted in INSPIRE as of 17 Aug 2014
  C.~Germani and Y.~Watanabe,
  %``UV-protected (Natural) Inflation: Primordial Fluctuations and non-Gaussian Features,''
  JCAP {\bf 1107}, 031 (2011)
  [Addendum-ibid.\  {\bf 1107}, A01 (2011)]
  [arXiv:1106.0502 [astro-ph.CO]].
  %%CITATION = ARXIV:1106.0502;%%
  %42 citations counted in INSPIRE as of 17 Aug 2014
  
  
  
%\cite{Dimopoulos:2005ac}
\bibitem{Dimopoulos:2005ac} 
  S.~Dimopoulos, S.~Kachru, J.~McGreevy and J.~G.~Wacker,
  %``N-flation,''
  JCAP {\bf 0808}, 003 (2008)
  [hep-th/0507205].
  %%CITATION = HEP-TH/0507205;%%
  %277 citations counted in INSPIRE as of 22 Apr 2014

%\cite{Liddle:1998jc}
\bibitem{Liddle:1998jc} 
  A.~R.~Liddle, A.~Mazumdar and F.~E.~Schunck,
  %``Assisted inflation,''
  Phys.\ Rev.\ D {\bf 58}, 061301 (1998)
  [astro-ph/9804177];
  E.~J.~Copeland, A.~Mazumdar and N.~J.~Nunes,
  %``Generalized assisted inflation,''
  Phys.\ Rev.\ D {\bf 60}, 083506 (1999)
  [astro-ph/9904309];
  P.~Kanti and K.~A.~Olive,
  %``Assisted chaotic inflation in higher dimensional theories,''
  Phys.\ Lett.\ B {\bf 464}, 192 (1999)
  [hep-ph/9906331].
  %%CITATION = HEP-PH/9906331;%%
  %99 citations counted in INSPIRE as of 22 Apr 2014

%\cite{Silverstein:2008sg}
\bibitem{Silverstein:2008sg} 
  E.~Silverstein and A.~Westphal,
  %``Monodromy in the CMB: Gravity Waves and String Inflation,''
  Phys.\ Rev.\ D {\bf 78}, 106003 (2008)
  [arXiv:0803.3085 [hep-th]];
  L.~McAllister, E.~Silverstein and A.~Westphal,
  %``Gravity Waves and Linear Inflation from Axion Monodromy,''
  Phys.\ Rev.\ D {\bf 82}, 046003 (2010)
  [arXiv:0808.0706 [hep-th]].
  %%CITATION = ARXIV:0808.0706;%%
  %199 citations counted in INSPIRE as of 24 Apr 2014  

%\cite{Marchesano:2014mla}
\bibitem{Marchesano:2014mla} 
  F.~Marchesano, G.~Shiu and A.~M.~Uranga,
  %``F-term Axion Monodromy Inflation,''
  arXiv:1404.3040 [hep-th];
  A.~Hebecker, S.~C.~Kraus and L.~T.~Witkowski,
  %``D7-Brane Chaotic Inflation,''
  arXiv:1404.3711 [hep-th].


  %%CITATION = ARXIV:1207.1128;%%
  %9 citations counted in INSPIRE as of 26 Apr 2014

%\cite{Chatzistavrakidis:2012bb}

%\cite{Harigaya:2014eta}


%\cite{Kaloper:2008fb}
\bibitem{Kaloper:2008fb} 
  N.~Kaloper and L.~Sorbo,
  %``A Natural Framework for Chaotic Inflation,''
  Phys.\ Rev.\ Lett.\  {\bf 102}, 121301 (2009)
  [arXiv:0811.1989 [hep-th]];
  N.~Kaloper, A.~Lawrence and L.~Sorbo,
  %``An Ignoble Approach to Large Field Inflation,''
  JCAP {\bf 1103}, 023 (2011)
  [arXiv:1101.0026 [hep-th]].
  %%CITATION = ARXIV:1101.0026;%%
  %59 citations counted in INSPIRE as of 06 May 2014
  \bibitem{Harigaya:2014eta}
 K.~Harigaya and M.~Ibe,
 %``Inflaton potential on a Riemann surface,''
 arXiv:1404.3511 [hep-ph].



  
\bibitem{Tye:2014tja} 
  S.~-H.~H.~Tye and S.~S.~C.~Wong,
  %``Helical Inflation and Cosmic Strings,''
  arXiv:1404.6988 [astro-ph.CO]
  
\bibitem{Ben-Dayan:2014zsa} 
  I.~Ben-Dayan, F.~G.~Pedro and A.~Westphal,
  %``Hierarchical Axion Inflation,''
  arXiv:1404.7773 [hep-th].\  
  
  \bibitem{Chatzistavrakidis:2012bb} 
  A.~Chatzistavrakidis, E.~Erfani, H.~P.~Nilles and I.~Zavala,
  %``Axiology,''
  JCAP {\bf 1209}, 006 (2012)
  [arXiv:1207.1128 [hep-ph]].

\end{thebibliography}
\end{document}